\documentclass[reprint,amsmath,amssymb,aps,showkeys]{revtex4-2}

\usepackage{float}
\usepackage{caption}
\usepackage{subcaption}
\usepackage{graphicx}
\usepackage{dcolumn}
\usepackage{bm}

\begin{document}
	
\title{Relativistic effects in the study of structure and electronic properties of UO$_2$ \\ within DFT+U method}
\author{Mahmoud Payami }
\email{mpayami@aeoi.org.ir}
\author{Samira Sheykhi}
\affiliation{School of Physics \& Accelerators, Nuclear Science and Technology Research Institute, AEOI,\\ 
	P.~O.~Box~14395-836, Tehran, Iran}

\begin{abstract}
To study crystals that contain heavy atoms, it is important to consider the relativistic effects, as electrons in orbitals close to the atom's nucleus can reach speeds comparable to that of light in a vacuum. In this study, we utilized the first-principles DFT+U method to analyze the electronic structure and geometric properties of uranium dioxide (UO2) using three formulations: full-relativistic, scalar-relativistic, and non-relativistic. Our findings demonstrate that the non-relativistic scheme produces results that deviate significantly from experimental values for both lattice constant and band gap. In contrast, the scalar-relativistic regime yields highly accurate results for the geometric properties of UO2, and is therefore sufficient for most studies. However, for a more precise analysis, the full-relativistic calculations with spin-orbit effects should be employed, which result in a $6.2\%$ increase in the Kohn-Sham band-gap and a $0.05\%$ decrease in the lattice constant compared to the scalar-relativistic approach.
\end{abstract}

\keywords{Uranium dioxide; Anti-ferromagnetism; Density-Functional Theory; Spin-orbit effect; Mott Insulator; DFT+U.}

\maketitle

\section{Introduction}\label{sec1}

\begin{figure}[ht]
	\centering
	\includegraphics[width=0.3\linewidth]{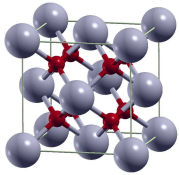}
	\vspace*{8pt}
	\caption{UO$_2$ crystal structure at low temperatures with cubic space group $Fm\bar{3}m$ (No. 225) with lattice constant of 5.47$\AA$.} \label{fig1}
\end{figure}

UO$_2$ is frequently used as a fuel in nuclear power reactors. Experimental investigations have revealed that UO$_2$ possesses an anti-ferromagnetic (AFM) crystal structure with a 3k-order at temperatures below $30~$K. Conversely, at higher temperatures, UO$_2$ takes on a para-magnetic form.\cite{Amoretti,Faber} 
Previous experimental findings \cite{idiri2004behavior} demonstrated that uranium and oxygen atoms occupy the octahedral (4a) and tetrahedral (8c) symmetry positions, respectively, within the cubic space group $Fm\bar{3}m$ (No. 225) with a lattice constant of 5.47\AA, as depicted in Fig.~\ref{fig1}. However, recent XRD experiments \cite{desgranges2017actual} have revealed that UO$_2$ crystallizes with a less symmetric cubic space group $Pa\bar{3}$ (No. 205), with oxygen atoms slightly displaced inside the cube.

Previous research has investigated the electronic structure of UO$_2$ \cite{baer1980electronic,schoenes1978optical,gubanov1977electronic,dudarev1998electronic,schoenes1980electronic,dudarev1997effect,dorado2009dft+,pegg2017dft+,SHEYKHI201893,christian2021interplay}. It is widely recognized that the standard approximations used in density-functional theory (DFT) \cite{hohenberg1964,kohn1965self} to describe the system often result in incorrect metallic behavior, whereas experimental observations indicate that UO$_2$ is an insulator, a phenomenon referred to as the "Mott insulator." This incorrect metallic prediction is due to the standard approximations treating the partially-filled "localized" $5f$ and/or $6d$ valence electrons in uranium atoms, or $2p$ valence electrons in oxygen atoms, on par with other "delocalized" electrons. To address this issue, researchers commonly employ the DFT+U method \cite{cococcioni2005linear,himmetoglu2014hubbard,dorado2009dft+,freyss4}, which is computationally less expensive and used in our calculations. Another approach involves utilizing orbital-dependent hybrid functionals for the exchange-correlation (XC) energy functional \cite{SHEYKHI201893}.

The high speeds of electrons in orbitals near the nuclei of heavy atoms, such as uranium, are well-known, and must be considered when analyzing their behavior. To estimate the speeds of these inner electrons in a uranium atom, we used a simple Bohr model for hydrogen, neglecting electron-electron interactions. The speed of the electron can be calculated using the formula $v_e=\sqrt{kZe^2/m_e r}$, where $k=8.99\times 10^9$, $m_e=9.11\times 10^{-31}$, and $e=1.6\times 10^{-19}$ in SI units. For a uranium atom with $Z=92$, we used the positions of the peak of atomic wave-functions as the radii of Bohr orbits and estimated the speeds of $1s$, $2p$, $3d$, and $4f$ electrons. The results are presented in Table~\ref{tab0}.
\begin{table}[ht]
	\caption{Estimation of the speed for inner electrons of U and O atoms using the simple Bohr model. $a_0=0.529\AA$ and $c$ is the speed of light. }
	\begin{center}    
		\begin{tabular}{ ccccc } 
			\hline \hline
			atom   &$Z$    & $nl$ & $r_e/a_0$ & $v_e/c$ \\ \hline 
			U      &92     & $1s$ & 0.008     &  0.80   \\ 
			&       & $2p$ & 0.041     &  0.40   \\
			&       & $3d$ & 0.114     &  0.20   \\
			&       & $4f$ & 0.286     &  0.10   \\   \hline
			O      & 8     & $1s$ & 0.129     &  0.06   \\   \hline   
		\end{tabular}\label{tab0}
	\end{center}
\end{table}
The data presented in Table~\ref{tab0} indicate that relativistic effects do not need to be considered for O atoms, whereas for U atoms, the speeds of the inner electrons are of a magnitude comparable to that of light.

The effects of relativity are typically accounted for using two levels of approximation: the "scalar-relativistic" approximation, where the spin-orbit interaction is negligible, and the "full-relativistic" approximation, where the spin-orbit effect is significant. In the full-relativistic approach, the Dirac-Kohn-Sham equations are solved.

In this study, we have utilized the DFT+U method to compute the electronic and geometric characteristics of uranium dioxide employing full-relativistic (FR), scalar-relativistic (SR), and non-relativistic (NR) formulations, and subsequently compared the outcomes. The uranium atoms were modeled utilizing the simplified 1k-order AFM configuration, and the constraint of $Fm\bar{3}m$ space group was applied during the geometry optimization stage, which had minimal impact on the results.
Our findings demonstrate that the equilibrium lattice constants in the FR and SR methods exhibit slight differences ($0.05\%$), while the electronic band gaps differ significantly ($6.2\%$). In contrast, the equilibrium lattice constants in the SR and NR cases show relatively large differences ($2\%$), and the gap energy difference is considerable ($64\%$). Therefore, for investigating UO$_2$, the SR approach is reasonably accurate as long as electronic excitation properties are not of concern.

This paper is organized as follows: Section~\ref{sec2} provides the computational details; Section~\ref{sec3} presents and discusses the calculated results; and Section~\ref{sec4} summarizes the conclusions of this study.

\section{Computational details}\label{sec2}
\subsection{Pseudopotentials}
The norm-conserving pseudo-potentials (NCPP) used for U and O atoms were generated by the {\it APE} code, and to ensure transferability, nonlinear core correction was applied during the generation step for all three cases (NR, SR, and FR). The reference valence configurations of U($6s^2$, $6p^6$, $7s^2$, $7p^0$, $6d^1$, $5f^3$) and O($2s^2$, $2p^4$) were used in the NCPP generation for all three cases. In the FR case, the pseudo-potentials were generated by solving Dirac's equation, as described in Appendix~\ref{appendix}. 

 In the NR case, the KS equations were solved self-consistently in the spherically symmetric effective potential for the atom. The simplified Dirac's equations were used in the SR case, in which the spin-orbit term in the Hamiltonian was initially omitted but the "mass-velocity" and "Darwin" terms were retained \cite{koelling1977technique}. For the FR case, the two-coupled equations (\ref{eq4a})-(\ref{eq4b}) were solved self-consistently. It should be noted that even the valence electrons, which may have small speeds compared to the inner relativistic electrons, undergo modifications in their orbitals due to the modified effective potentials resulting from the contributions of high-speed core electrons.

\subsection{Electronic structure of UO$_2$}
All calculations presented in this paper are based on the solution of the KS equations in DFT, utilizing the Quantum-ESPRESSO code package. \cite{Giannozzi_2009,doi:10.1063/5.0005082}
To ensure the accuracy of our calculations, we performed convergence tests and determined the appropriate kinetic energy cutoffs for the plane-wave expansions to be 350 and 1400Ry for the wave functions and charge densities, respectively. To avoid self-consistency issues, we used the Methfessel-Paxton smearing method \cite{methfessel1989high} with a width of 0.01Ry for the occupations.
For geometry optimizations, we employed a $6\times 6\times 6$ grid with a shift for Brillouin-zone integrations. For density-of-states (DOS) calculations, we utilized a denser grid of $8\times 8\times 8$ in reciprocal space and the "tetrahedron" method \cite{PhysRevB.49.16223} for the occupations.
In DFT+U calculations, we used a Hubbard-U parameter value of 4.0 eV, consistent with values determined by other works.\cite{yamazaki1991systematic,kotani1992systematic} We applied the $Fm\bar{3}m$ constraint to optimize all geometries for total pressures on unit cells to within 0.5 kbar, and forces on atoms to within 10$^{-6}$~Ry/a.u.
To handle the multi-minima total energy function for the lowest energy in DFT+U approach, we utilized the occupation-matrix control (OMC) method, previously used by others.\cite{dorado2009dft+,freyss4}
After examining different XC schemes, we found that the generalized gradient approximation (GGA-PBEsol)\cite{PhysRevLett.100.136406,PhysRevLett.102.039902} produced the best agreement with experimental lattice constant and band-gap values. Hence, we employed GGA-PBEsol for our calculations. Finally, we used the simplified model of 1k-order AFM configuration for uranium atoms in all our calculations.

\section{Results and discussions}\label{sec3}
\subsection{Pseudopotentials}
\begin{figure*}[ht]
	\centering
	\includegraphics[width=0.97\linewidth]{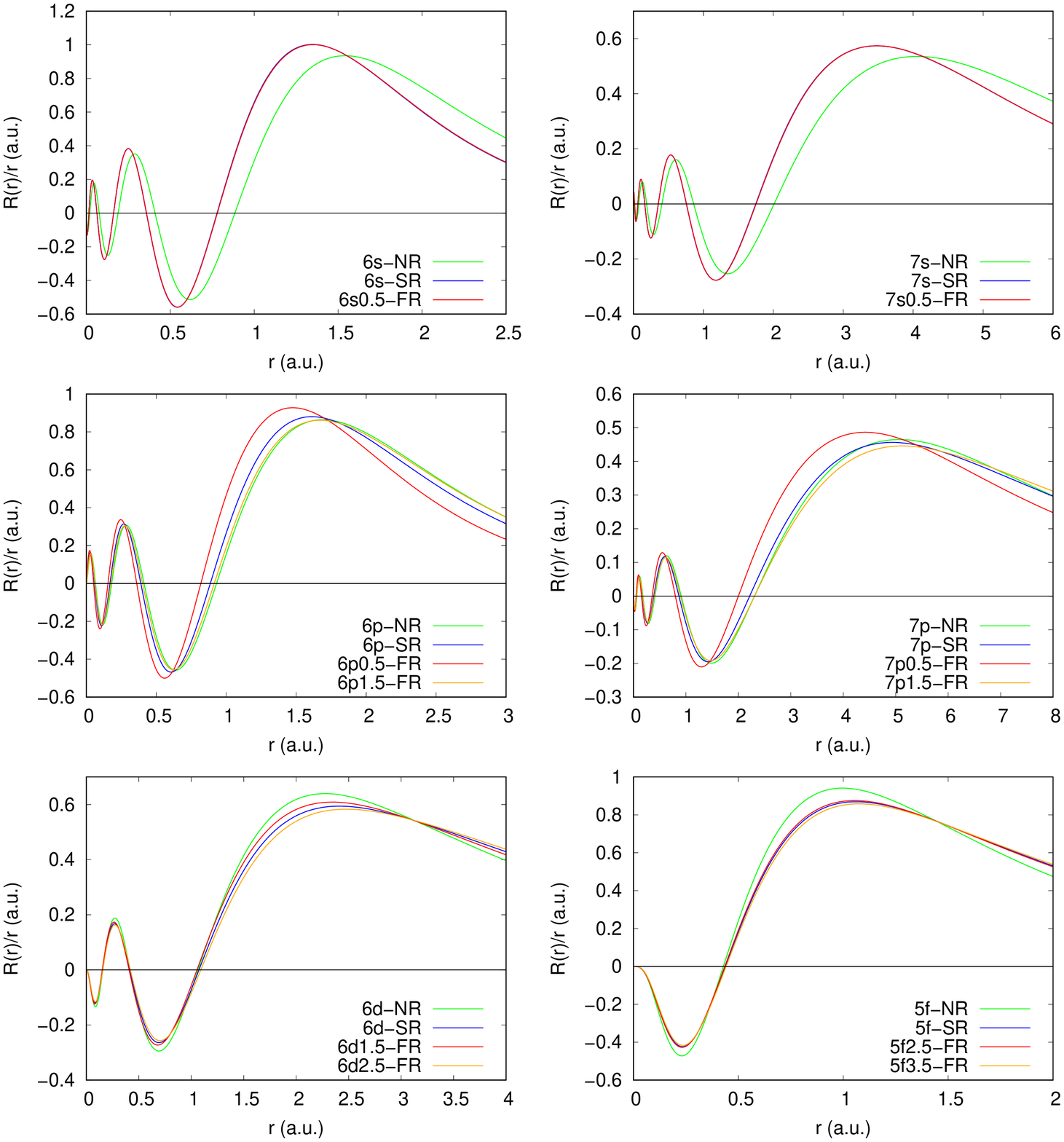}
	\vspace*{8pt}
	\caption{All-electron U-atomic valence orbitals for NR, SR, and FR cases. As is seen, the $6s$ and $7s$ orbitals in the NR case are spatially expanded relative to SR and FR cases, and this, in turn, leads to the increase of equilibrium lattice constant in NR level of computations.} 
	\label{fig2}
\end{figure*}

\begin{figure*}[ht]
	\centering
	\includegraphics[width=0.97\linewidth]{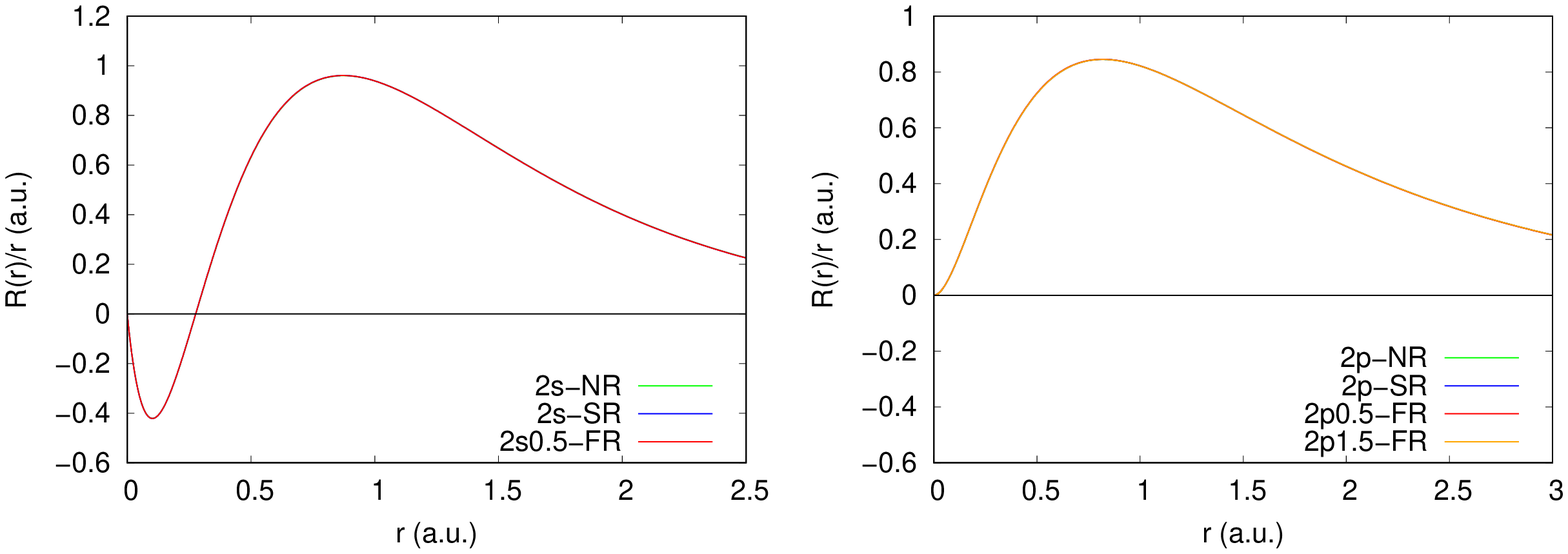}
    \vspace*{8pt}
	\caption{All-electron O-atomic valence orbitals for NR, SR, and FR cases. As is seen, the $2s$ and $2p$ orbitals in all cases of NR, SR, and FR are more or less the same.}
	\label{fig3}
\end{figure*}

The pseudo-potentials used for the U and O atoms were generated using the APE code. In Figure~\ref{fig2}, we compare the radial wave functions of the atomic all-electron valence orbitals ($6s$, $6p$, $7s$, $7p$, $6d$, and $5f$) for the U atom. The $6s$ and $7s$ orbitals in the NR case are expanded spatially compared to the SR and FR cases, leading to a larger equilibrium lattice constant in the NR case. Figure~\ref{fig3} shows the comparison of the atomic all-electron valence orbitals ($2s$ and $2p$) for O. The relativistic treatment of the O atom does not cause any corrections to the NR orbitals, but we use the same regime for the U and O pseudo-potentials in NR, SR, and FR calculations of UO$_2$ for consistency.

\subsection{Electronic structure of UO$_2$}
As previously stated, the GGA-PBEsol approximation for exchange-correlation yields excellent agreement with experimental results. Orthonormalized projection on to Hubbard orbitals was used. To account for the anti-ferromagnetic configuration of U atoms, a simplified 1k-AFM configuration was utilized, in which the U atoms alternately change magnetization in the $z$ direction. Geometries were optimized with the $Fm\bar{3}m$ space group constraint. To prevent meta-stable states in the DFT+U calculations, the OMC method was employed.

Table~\ref{tab1} displays the equilibrium lattice constants for each of the three regimes considered. The results indicate that the difference in lattice constant between the FR and SR approaches is very small ($0.05\%$), whereas the electronic gaps differ significantly (by $6.2\%$). In contrast, the equilibrium lattice constants in the SR and NR cases differ relatively significantly (by $2\%$), while the difference in the energy gap is very large ($64\%$). These findings suggest that the NR scheme for both geometric and electronic properties is far from experimental values and should not be used in the study of UO$_2$. However, the SR regime is quite accurate when considering geometric properties, while the FR regime provides better accuracy when studying electronic excitation properties.

\begin{table}[ht]
	\caption{Equilibrium lattice constant, $a$, in \AA; Kohn-Sham electronic band gap, $E_g$, in eV for the three schemes of NR, SR, and FR compared with experimental values. }
	\begin{center}    
		\begin{tabular}{ ccc } 
			\hline \hline
			scheme & $a (\AA)$ & $E_g$ (eV) \\ \hline 
			NR     & 5.588     & 0.97     \\ 
			SR     & 5.480     & 2.72     \\
			FR     & 5.477     & 2.90     \\
			Exp.   & 5.470     & 2.20     \\   \hline
		\end{tabular}\label{tab1}
	\end{center}
\end{table}

\begin{figure*}[ht]
	\centering
	\includegraphics[width=0.5\linewidth]{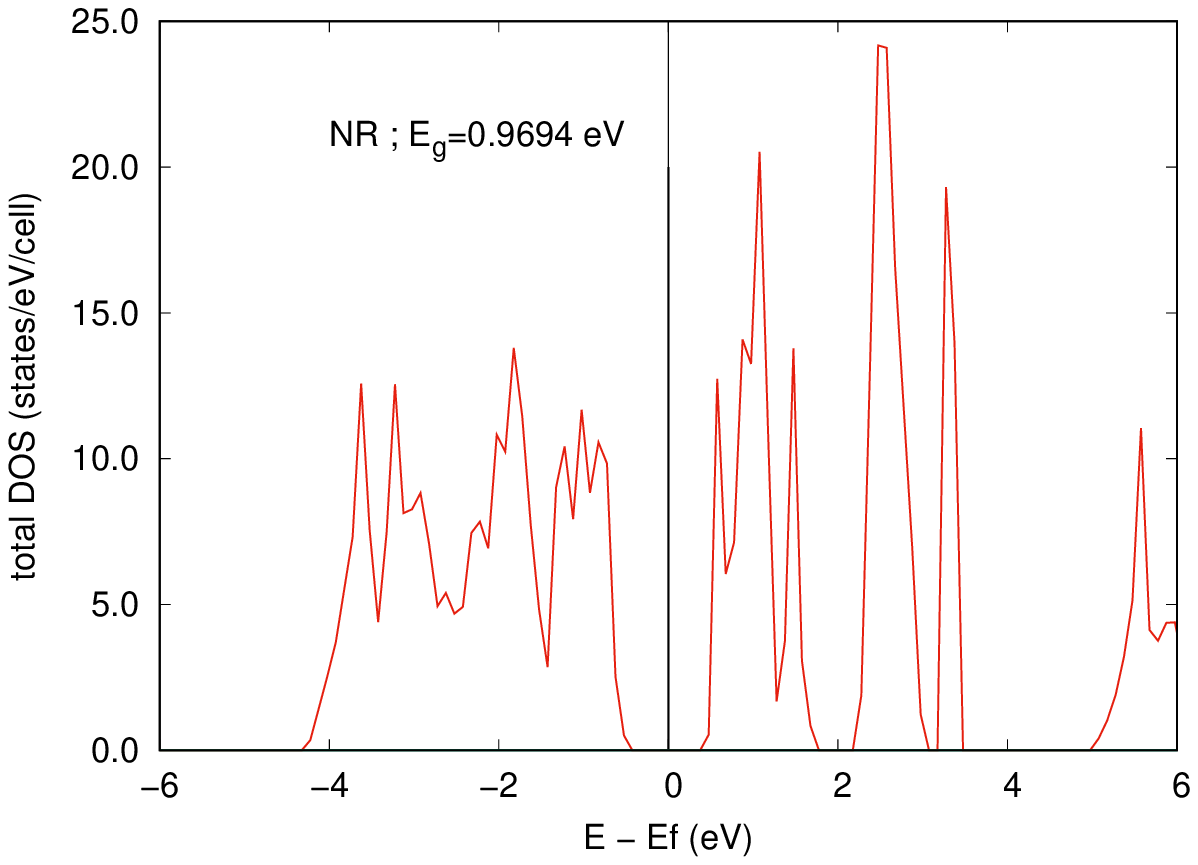}
	\centering
	\includegraphics[width=0.5\linewidth]{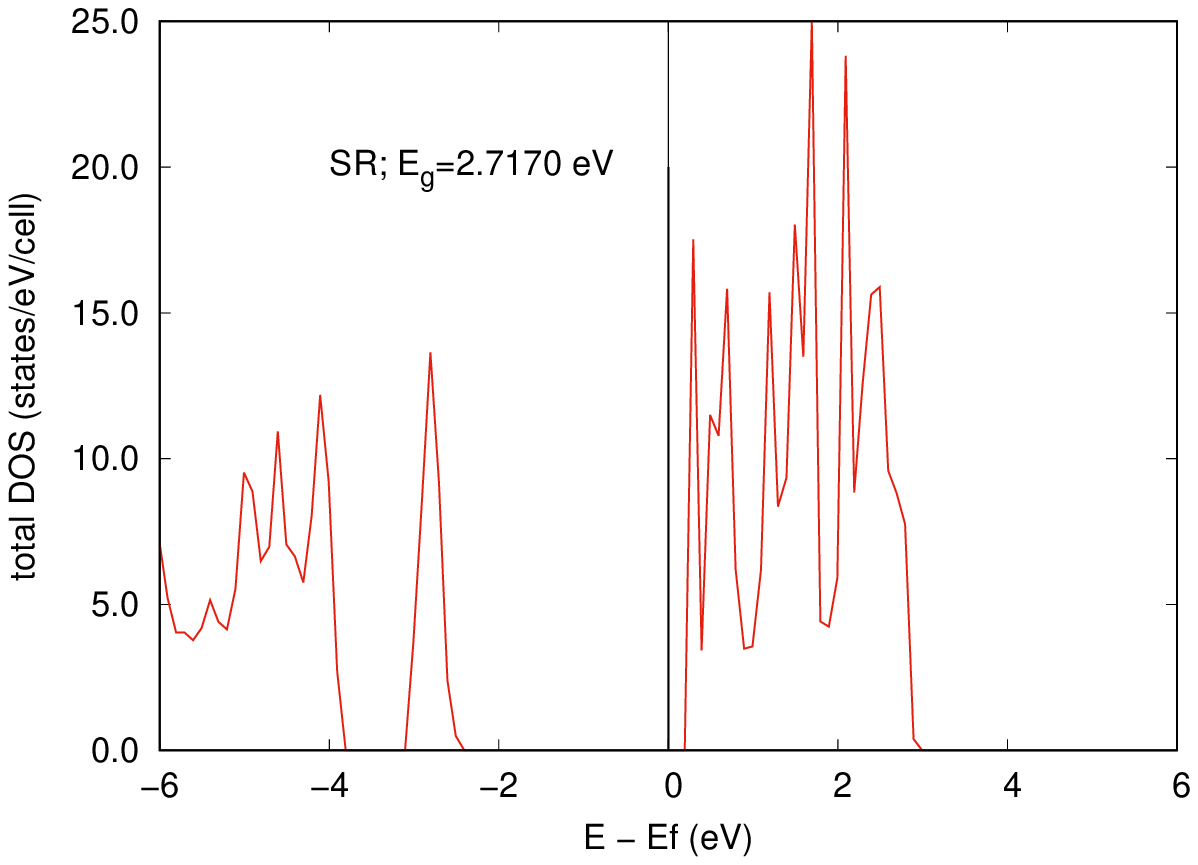}
	\centering
	\includegraphics[width=0.5\linewidth]{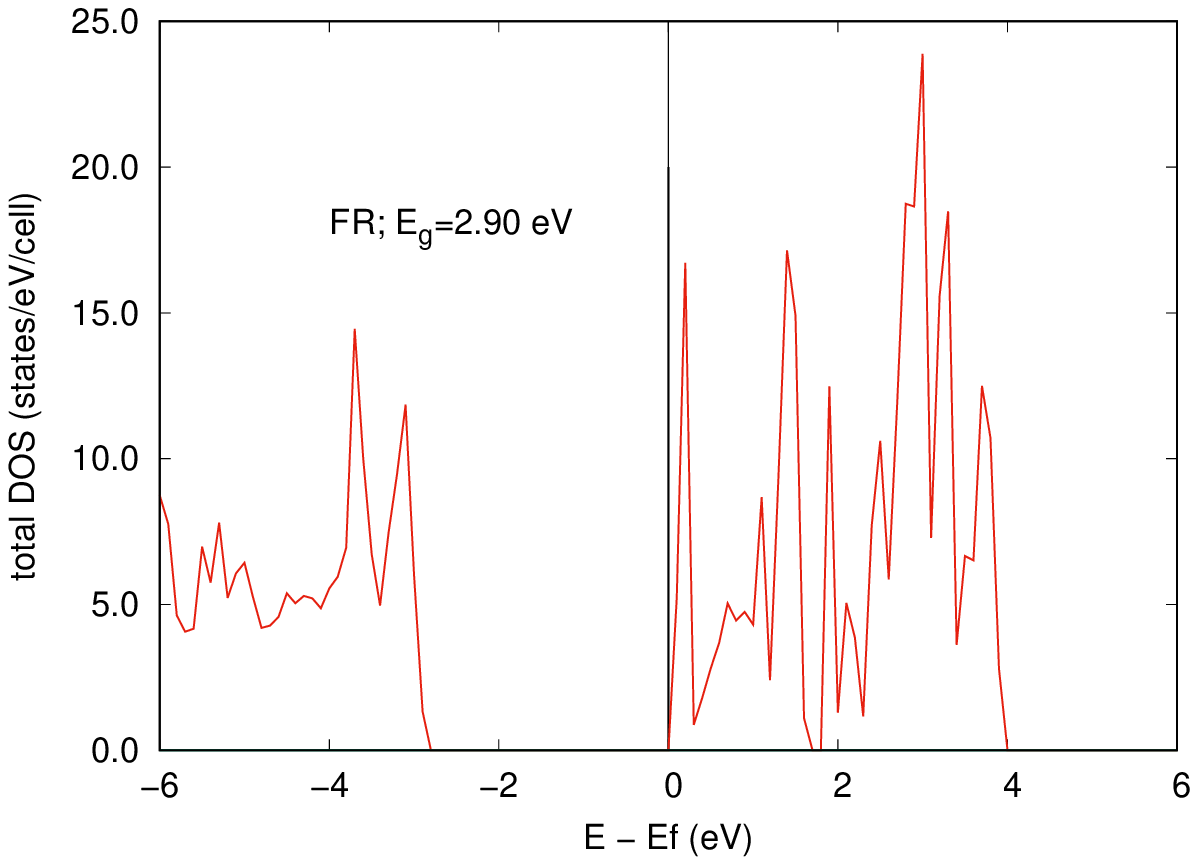}
    \vspace*{8pt}
    \caption{Electron density of states (DOS) for the NR, SR, and FR cases. All the three cases are predicted to be insulators.}
    \label{fig4}
\end{figure*}

Fig.~\ref{fig4} illustrates the total electron density of states (DOS) for the three scenarios. It can be observed that all three ground states are insulators.

\section{Conclusions}\label{sec4}
This study demonstrated the necessity of considering relativistic effects on the inner electrons of U atoms, while such effects are negligible for O atoms. To investigate the properties of UO$_2$ crystal, NCPPs were generated at three levels of NR, SR, and FR. The U atoms' AFM configuration was modeled using a simplified $1k$-order AFM. To prevent meta-stable states in the DFT+U calculations, the occupation-matrix control method was employed. The results indicated that the NR scheme is not reliable for both geometric and electronic properties, and thus, should be avoided. Conversely, the SR and FR approximations produce comparable and accurate geometric properties, with only a $0.05\%$  difference in lattice constant. 
Nevertheless, the band gaps vary by $6.2\%$, leading to differences in the excitation energies.   

\section*{Acknowledgement}
This work is part of research program in School of Physics and Accelerators, NSTRI, AEOI.  

\section*{Data availability }
The raw or processed data required to reproduce these results can be shared with anybody interested upon 
sending an email to M. Payami.

\appendix
\section{Dirac Equation}\label{appendix}

The Dirac equation for an electron in a scalar potential $V$ is given by
\begin{equation}
	\hat{H}_D\Psi(\vec{r})=E\Psi(\vec{r}),
\end{equation}

\begin{equation}
	\hat{H}_D = c ( \vec{\alpha} \cdot \vec{p}) + \beta m c^2 + VI_4,
\end{equation}

\begin{equation}
	\alpha_i = 
	\begin{pmatrix}
		0 & \sigma_i \\
		\sigma_i & 0
	\end{pmatrix}
	\;\;\;\;\;
	\beta = 
	\begin{pmatrix}
		I_2 & 0 \\
		0 & I_2
	\end{pmatrix}
	\;\;\; ;\;\;\;(i=1,2,3),
\end{equation}
in which $\sigma_i$'s are Pauli matrices and $I_2$ is a two-dimensional unit matrix. $\vec{p}$, $m$, and $c$ are electron momentum, electron rest mass, and light speed in vacuum, respectively.

For a scalar potential with spherical symmetry, it can be shown that the Dirac equation transforms to the following two coupled equations:
\begin{equation}
	\begin{split}
		-\frac{\hbar^2}{2M} \frac{1}{r^2} \frac{d}{dr} \left[ r^2 \frac{dg_{n\kappa}}{dr}\right] + \left[ V+\frac{\hbar^2}{2M}\frac{l(l+1)}{r^2}\right] g_{n\kappa} \\
		-\frac{\hbar^2}{4M^2 c^2} \frac{dV}{dr} \frac{dg_{n\kappa}}{dr} \;\;\;\;\;\;\;\;\;\;\;\;\;\;\;\;\;\;\;\;\;\;\;\;\;\;\; \\
		-\frac{\hbar^2}{4M^2 c^2} \frac{dV}{dr} \frac{1+\kappa}{r} g_{n\kappa} = E^\prime g_{n\kappa}
	\end{split} \label{eq4a}
\end{equation}

\begin{equation}
	\frac{df_{n\kappa}}{dr} = \frac{1}{\hbar c}\left( V-E^\prime \right) g_{n\kappa} + \frac{\kappa -1}{r} f_{n\kappa}\label{eq4b}
\end{equation}
Here, we have used

\begin{equation}
	\Psi(\vec{r})=
	\begin{pmatrix}
		\psi^1 (\vec{r}) \\
		\psi^2 (\vec{r})
	\end{pmatrix}
	=
	\begin{pmatrix}
		g_{nj}(r)\phi_{jlm}(\theta , \phi) \\
		if_{nj}(r) \phi_{jlm}(\theta , \phi)
	\end{pmatrix}, \label{eq5}
\end{equation}
where the spinor-angle functions $\phi_{jlm}$ are defined as products of spherical harmonics and spinors, and for $j=l\pm 1/2$ are given by:

\begin{equation}
	\begin{split} 
	\phi_{jlm}=\left[ \frac{(l+1/2)\pm m}{2l+1}\right]^{1/2} Y_l^{m-1/2} 
	\begin{pmatrix}
		1 \\
		0   
	\end{pmatrix} \\
	\pm
	\left[ \frac{(l+1/2)\mp m}{2l+1}\right]^{1/2} Y_l^{m+1/2} 
	\begin{pmatrix}
		0 \\
		1   
	\end{pmatrix} \label{eq6}
    \end{split}
\end{equation} 

In the above equations, the following definitions were used:

\begin{equation}
	\begin{split}
		E^\prime = E-mc^2, \;\; M(r)=m+\frac{E^\prime - V(r)}{2c^2} \\
		j=l+\frac{\lambda}{2}' \;\;\; \kappa=-\lambda(j+\frac{1}{2}), \;\; \lambda=+1,-1 \\
		j=l+\frac{1}{2} \rightarrow \kappa=-(l+1), \;\;\; j=l-\frac{1}{2} \rightarrow \kappa=+l
	\end{split} \label{eq7}
\end{equation}
In the left hand side of Eq.~(\ref{eq4a}), the second, third, and fourth terms are the so-called "mass-velocity", "Darwin", and "spin-orbit" terms.


\vspace*{2cm}
\section*{References}

\begin{thebibliography}{0}
	
\bibitem{Amoretti} G. Amoretti, A. Blaise, R. Caciuffo, et. al., {\it Phys. Rev.} {\bf B40}, 1856 (1989).

\bibitem{Faber} J. Faber, G. H. Lander, and B. R. Cooper, {\it Phys. Rev. Lett.} {\bf 35}, 1770 (1975).

\bibitem{idiri2004behavior} M. Idiri, T. Le Bihan, S. Heathman, et. al., {\it Phys. Rev.} {\bf B70}, 014113 (2004).

\bibitem{desgranges2017actual} L. Desgranges, Y. Ma, Ph. Garcia, et. al., {\it Inorg. Chem.} {\bf 56}, 321 (2017).

\bibitem{baer1980electronic} Y. Baer and J. Schoenes, {\it Solid State Commun.} {\bf 33}, 885 (1980).

\bibitem{schoenes1978optical} J. Schoenes, {\it J. Appl. Phys.} {\bf 49}, 1463 (1978).

\bibitem{gubanov1977electronic} V. A. Gubanov, A. Rosen, and D. E. Ellis, {\it Solid State Commun.} {\bf 22}, 219 (1977).

\bibitem{dudarev1998electronic} S. L. Dudarev, G. A. Botton, S. Yu Savrasov, et. al., {\it Phys. Stat. Sol. (a)} {\bf 166}, 429 (1998).

\bibitem{schoenes1980electronic} J. Schoenes, {\it Phys. Rep.} {\bf 63}, 301 (1980).

\bibitem{dudarev1997effect} S. L. Dudarev, D. Nguyen Manh, and A. P. Sutton, {\it Phil. Mag.} {\bf B75}, 613 (1997).

\bibitem{dorado2009dft+} B. Dorado, B. Amadon, M. Freyss, et. al., {\it Phys. Rev.} {\bf B79}, 235125 (2009).

\bibitem{pegg2017dft+} J. T. Pegg, X. Aparicio-Angles, M. Storr, et. al., {\it J. Nucl. Mat.} {\bf 492}, 269 (2017).

\bibitem{SHEYKHI201893} S. Sheykhi and M. Payami, {\it Physica C: Superconductivity and its Applications} {\bf 549}, 93 (2018).

\bibitem{christian2021interplay} M. S. Christian, E. R. Johnson, and T. M. Besmann, {\it J. Phys. Chem.} {\bf A125}, 2791 (2021).

\bibitem{hohenberg1964} P. Hohenberg and W. Kohn, {\it Phys. Rev.} {\bf 136}, B864 (1964).

\bibitem{kohn1965self} W. Kohn and L. J. Sham, {\it Phys. Rev.} {\bf 140}, A1133 (1965).

\bibitem{cococcioni2005linear} M. Cococcioni and S. De Gironcoli, {\it Phys. Rev.} {\bf B71}, 035105 (2005).

\bibitem{himmetoglu2014hubbard} B. Himmetoglu, A. Floris, S. De Gironcoli, et. al., {\it Int. J. Quantum Chem.} {\bf 114}, 14 (2014).

\bibitem{freyss4} M. Freyss, B. Dorado, M. Bertolus, et. al., {\it $\psi_k$~Scientific~Highlight~Of~The~Month} No. 113 (2012).

\bibitem{OLIVEIRA2008524} M. J. T. Oliveira and F. Nogueira, {\it Comput. Phys. Commun.} {\bf 178}, 524 (2008).

\bibitem{koelling1977technique} D. D. Koelling and B. N. Harmon, {\it J. Phys. C: Solid State Phys.} {\bf 10}, 3107 (1977).

\bibitem{Giannozzi_2009} P. Giannozzi, S. Baroni, N. Bonini, et. al., {\it J. Phys.: Condensed Matt.} {\bf 21}, 395502 (2009).

\bibitem{doi:10.1063/5.0005082} P. Giannozzi, O. Baseggio, P. Bonfà, et. al., {\it J. Chem. Phys.} {\bf 152}, 154105 (2020).

\bibitem{methfessel1989high} M. Methfessel and A. T. Paxton, {\it Phys. Rev.} {\bf B40}, 3616 (1989).

\bibitem{PhysRevB.49.16223} P. E. Bl\"ochl, O. Jepsen, and O. K. Andersen, {\it Phys. Rev.} {\bf B49}, 16223 (1994).

\bibitem{yamazaki1991systematic} T. Yamazaki and A. Kotani, {\it J. Phys. Soc. Japan} {\bf 60}, 49 (1991).

\bibitem{kotani1992systematic} A. Kotani and T. Yamazaki, {\it Prog. Theor. Phys. Suppl.} {\bf 108}, 117 (1992).

\bibitem{PhysRevLett.100.136406} J. P. Perdew, A. Ruzsinszky, G. Csonka, et. al., {\it Phys. Rev. Lett.} {\bf 100}, 136406 (2008).

\bibitem{PhysRevLett.102.039902} J. P. Perdew, A. Ruzsinszky, G. Csonka, et. al., {\it Phys. Rev. Lett.} {\bf 102}, 039902 (2009).

\end{thebibliography}
\end{document}